# Best practices for the application of temperature- and illumination-dependent current density–voltage $J(V, T, i)$ and electron-beam induced current EBIC to novel thin film solar cells

Authors: Rupak Chakraborty,[1] Riley E. Brandt,[1] Vera Steinmann, Paul Rekemeyer, Silvija Gradečak, Tonio Buonassisi*

\* *Contact: buonassisi@mit.edu*
[1] *Co-first authors*
*Massachusetts Institute of Technology, Cambridge, MA 02139, USA*

**Abstract**

As the photovoltaic community accelerates the development of new absorber candidate materials towards high-performing PV devices, it is essential to follow best practices and leverage deeper characterization tools. We have identified temperature- and illumination-dependent current density–voltage $J(V, T, i)$ and electron-beam induced current (EBIC) measurements as two powerful PV device characterization techniques to evaluate the potential of novel absorber candidate materials. Herein, we focus on the experimental methods and best practices for applying $J(V, T, i)$ and EBIC, addressing particular challenges in sample preparation and mounting. We demonstrate these on the example of tin monosulfide, a promising PV absorber candidate material that shares characteristics of many novel thin-film PV absorbers – mechanically soft, polycrystalline, and used in heterojunction thin-film PV devices.

**0. Introduction**

The impressively rapid efficiency improvement but poor reliability of lead halide perovskite solar cells has re-invigorated the search for other promising thin film photovoltaic (PV) materials. "Perovskite-inspired materials" in particular define a class of materials that may achieve a similar level of defect tolerance to the lead halide perovskites, but with non-toxic and earth-abundant elements, as well as improved stability in air.[1,2]

To validate the PV potential of these new absorber candidates, there is an imperative to demonstrate high-performance PV devices. However, early-stage device performance can be affected by a large number of artifacts extrinsic to the absorber material, especially non-ideal contacts and shunting. There is a danger that low power conversion efficiencies (PCE) resulting from early-stage devices may be "false negatives" that dissuade further investigation into promising materials. Conversely, there is risk of incorrectly evaluating PV device efficiencies by underestimating temperature- or voltage-dependent effects on power conversion efficiencies. Advanced electrical characterization of devices, when performed properly, can yield critical insights into the performance-limiting mechanisms. Knowing what measurements to take, and how to take them, are crucial steps to elucidate the full potential of novel materials and to focus effort on the most impactful research directions to increase efficiency.

We have learned from the organic PV community, as well as in recent years from the perovskite PV community, that the race to report record efficiencies without proper care may damage a field's reputation.[3–6] As a response, guidelines on simple device characterization have been published, such as how to calibrate a solar simulator[7] and how to perform accurate current density–voltage *J(V)* characterization on novel organic and hybrid PV devices.[8–14]

For other early-stage PV materials, we may avoid similar pitfalls by ensuring that we follow best practices for emerging thin film PV device characterization. In previous work, we have identified several beneficial characterization techniques, such as temperature- and illumination-dependent current density–voltage *J*(*V, T, i*) and electron-beam induced current EBIC measurements, Detailed examples of *J*(*V, T, i*) and EBIC results on thin film solar cells are published elsewhere.[15,16] *J*(*V, T, i*) measurements allow the separation and identification of loss mechanisms by spatial region in a solar cell. Cross-sectional EBIC measurements visualize current pathways and recombination processes that contribute to current loss and/or device shunting in a solar cell.

Herein, we focus on the experimental methods and best practices for applying *J*(*V, T, i*) and EBIC measurements to emerging thin film PV devices. Since *J*(*V, T, i*) and EBIC are both electrical measurement techniques requiring a rectifying PV device architecture, we discuss different device architectures for thin film solar cells, the importance of controlling the active area of a solar cell, and device statistics in Section 1. Section 2 describes the experimental setup for *J*(*V, T, i*) measurements, highlighting challenges and providing experimental guidance. Section 3 focuses on device shunting as a common artifact masking thin film device performance. We present the use of cross-sectional EBIC to map current pathways within a device. The visualization of through-thickness current pathways can help to detect regions of low shunt resistance. We describe the experimental setup, sample preparation and provide best practices for EBIC map acquisition.

Here, we choose tin monosulfide (SnS) as a proof-of-concept material system as its mechanical and structural properties are very similar to other emerging thin film PV materials (*e.g.*, bismuth tri-iodide, antimony selenide, or the lead halide perovskites). The soft and polycrystalline nature of SnS thin films can make sample preparation and mounting challenging, which has motivated the development of new approaches discussed herein.

1. **Device architectures**

Traditionally, the development of novel absorber candidate materials into high-performing PV devices takes decades, as it requires not only the development of the absorber materials but also careful device engineering including the search for selective contacts.[17,18] Techniques such as *J*(*V, T, i*) and cross-sectional EBIC can help us identify solar cell performance losses and guide process and device optimization towards high power conversion efficiencies. Those techniques, however, require at minimum a simple heterojunction stack, comprising three or more layers, in which the absorber thin film is sandwiched between two selective contacts. In the simplest form, one selective contact is Ohmic, and the other is a Schottky or rectifying contact.

Figure 1 demonstrates two simple device architectures in substrate-style configuration (Figure 1a) and in superstrate-style configuration (Figure 1b). The substrate-style device is illuminated from the top, while the superstrate device is illuminated from the bottom through the substrate. (illumination is illustrated by the yellow arrows in Figure 1).

It is worth noting that the selective contact can comprise multiple layers such as in the example of a SnS solar cell in Figure 1c.[19,20] Here, a more resistive buffer layer is inserted between the SnS absorber bulk and the transparent conductive oxide to reduce the risk of device shunting and thus avoid "false negatives". Top metal electrodes are used to reduce the series resistance of the device. Practical challenges leading to device shunting will be further addressed in section 3.

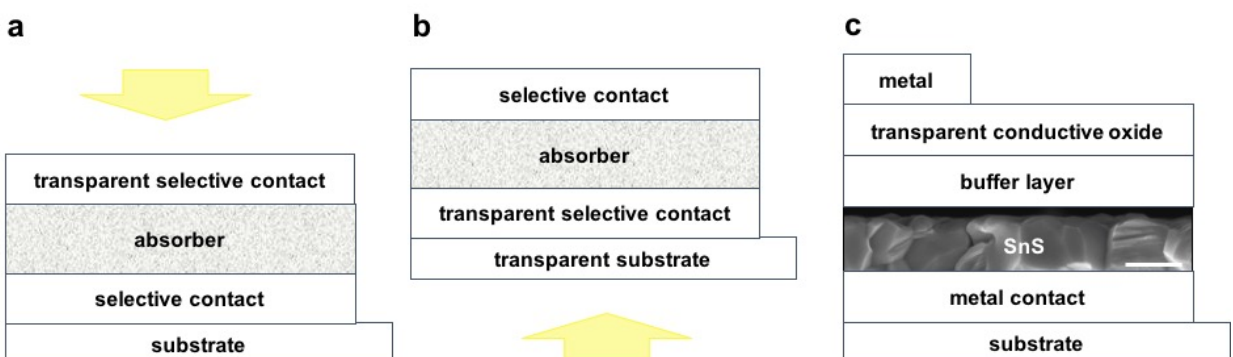

**Figure 1** Thin-film device architectures comprising a minimum of three layers: the absorber layer sandwiched between two selective contacts, forming an Ohmic contact on one side and a Schottky contact on the other. (a) Cross-sectional diagram of a general substrate-style device layer sequence. (b) Cross-sectional diagram of a general superstrate-style device layer sequence. The yellow arrows indicate the direction of illumination to operate the solar cell. (c) Example of a SnS substrate-style device architecture. The cross-sectional scanning electron micrograph shows the SnS absorber layer in the device stack; the scale bar indicates 500 nm.

### 1.1. Active area control

The active area of most laboratory-scale research cells is < 1 cm$^2$. At these small scales, edge effects erroneously affect the measured device performance because current is collected or recombining over a large area fraction outside of the nominal active device area. For established thin film PV technologies such as cadmium telluride (CdTe) and copper indium gallium selenide (CIGS), the active device area is commonly defined by a photolithographic device isolation process or mechanical scribing.[21,22] Those methods, however, can be time consuming to optimize and may not be compatible with certain chemistries or material softness (in the case of scribing). Photolithographic device isolation requires a series of steps (including the deposition, exposure and development of photoresist, followed by an etch step). Scribing isolation in principle only requires a sharp tool such as a razor blade that can cut through the layer stack outside the active device area. It can be challenging to achieve a "clean" scribe if the layered materials are either too hard or too soft. Many of the to-date investigated absorber candidates are rather soft materials, and scribing them leads to plastic deformation between layers and device shunting.

An alternative to the above methods is the use of a shadow mask, as shown in Figure 2. Here, the shadow mask is a rigid metal mask, which is used for defining the transparent contact area during deposition. It is also applied to the finished device when measuring the $J(V)$ device characteristics under illumination. To avoid shadowing effects during the illuminated $J(V)$ measurements, it is advised to attach the thin shadow mask directly to the device substrate by using tape or clips, for example. Thinner masks produce less shading, but are less rigid and may not make perfect contact with the substrate. Thicker masks can be improved by beveling the edges to reduce the shading effect, as seen in Figure 2.

In Figure 3, illuminated $J(V)$ measurements on SnS thin film devices are compared with and without a shadow mask. The measurements with and without a shadow mask are in good agreement within statistical error limits. However, we find that illuminated $J(V)$ measurements without a shadow mask systematically overestimate the device performance by 1-5% relative, suggesting that there is some charge carrier collection from outside the defined active area of the device.

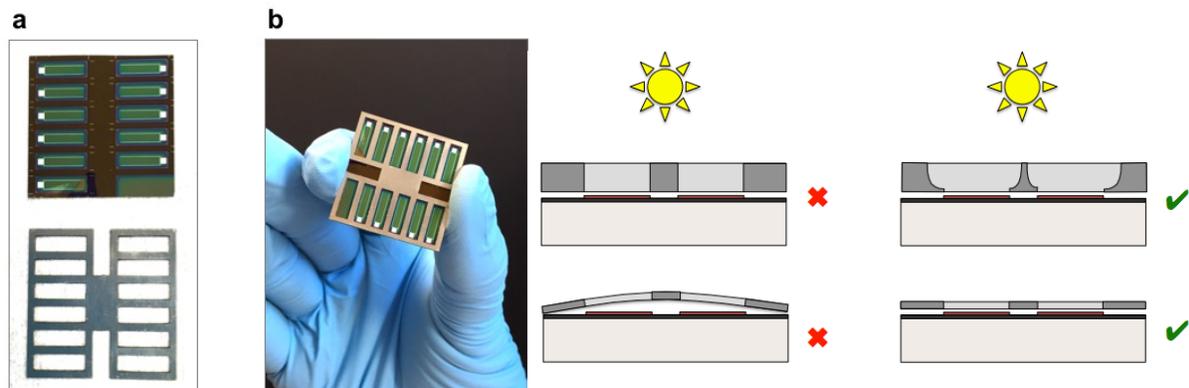

**Figure 2** Using a shadow mask to define the device area of SnS thin film devices. (a) Device substrates (2.5 x 2.5 $cm^2$) with eleven devices (each 0.25 $cm^2$) and the shadow mask. The device area is defined by the top contact deposition, using a shadow mask. The shadow mask is also applied during illuminated $J(V)$ measurements to avoid charge carrier collection from outside the device area. (b) Shadow mask placed on the device substrate prior to illuminated $J(V)$ measurements. The cross-sectional diagrams on the right demonstrate possible options of applying shadow masks onto a PV device for avoiding light shading or leakage effects.

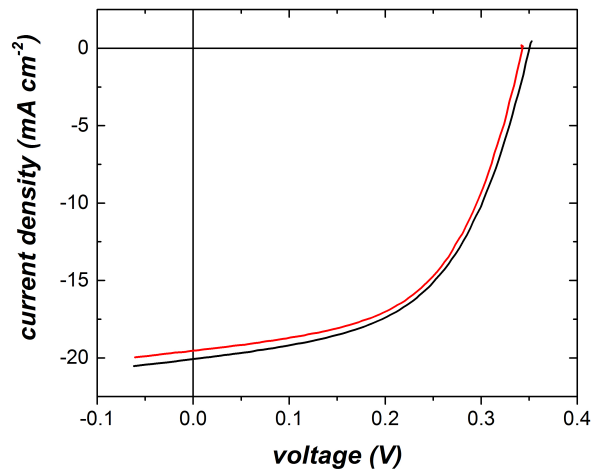

**Figure 3** Illuminated $J(V)$ measurements on a SnS solar cell without applying a shadow mask (black line) systematically overestimate device performance by 1-5% relative to those with shadow mask (red line).

## 1.2. Device statistics

Device statistics are necessary to evaluate the impact of changes in device fabrication on the resulting device performance.[23] Figure 2 shows a substrate with eleven identically fabricated solar cells on one substrate. Measuring all eleven devices reveals the spread in device performance due to both systematic and random spatial variation across one substrate. Repeating the fabrication and measurement process provides data on the reproducibility of the process, again with both systematic and random error contributions. Systematic errors should be characterized and minimized as best as possible using control samples for each fabrication run. In addition, repeating $J(V)$ measurements on all devices can identify measurement artifacts and quantify the relative contribution of errors due to measurement versus fabrication. The data is then analyzed as an ensemble to test hypotheses about the effect of process conditions on device performance using standard statistical practices, *i.e.*, computing sample averages, standard errors and confidence intervals. In general, reducing the process variance, which proportionally reduces standard error, is of primary importance in avoiding inconclusive hypothesis testing. Increasing the sample size for each process also reduces the standard error, though with less of an influence. As one example, one of our previous publications on SnS substrate-style devices addresses the challenge of improving performance reproducibility via process engineering and discusses solar cell ensemble data analysis in more detail.[23]

## 2. Temperature- and Illumination-dependent $J(V, T, i)$ measurements

Performing electrical characterization as a function of temperature and light intensity can help identify the limiting recombination mechanism,[16,24] remove artifacts associated with series resistance,[25,26] and help extract materials properties such as a heterojunction band offset.[16] However, it can be a time-intensive measurement and is prone to a number of artifacts, which easily influence the results.

$J(V, T, i)$ measurements are performed in a cryostat which enables substrate cooling below 0°C (avoiding condensation from ambient air), and using a solar simulator lamp to produce a one-sun spectrum as the light source. Alternatively, a temperature-controlled stage using heaters or thermoelectrics may be used for measurements in the range of 0–100°C, and other light sources such as monochromatic LEDs may be substituted. The following sections discuss best practices for cryostat-based measurements.

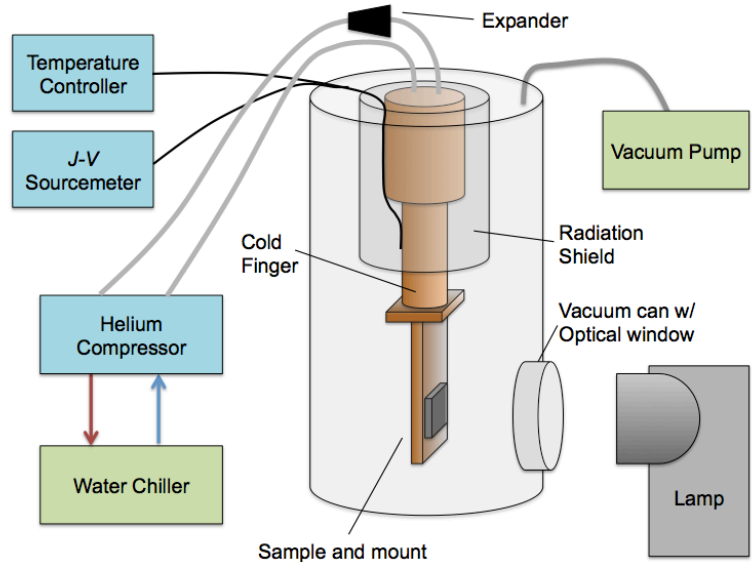

**Figure 4** Cryostat components for performing *J(V, T, i)* measurements. The copper cold finger, onto which the sample mount attaches, reaches temperatures as low as 4 K and as high as 340 K, controlled by the temperature controller. The compressor and expander control the refrigeration cycle for the helium and exchanges heat with a chilled water loop.[16]

## 2.1. Cryostat components

To cool the sample, a helium closed-loop cryostat cold finger is used to cover a temperature range from 4 K to 340 K. Cooling is provided by a helium refrigeration cycle, including an expander inside the cold finger, and an external compressor, which recompresses the helium and transfers the heat to chilled water in a heat exchanger. This water then exchanges heat in a water chiller. The cryostat components as part of the cooling cycle are shown in Figure 4.

Measuring the sample temperature accurately is critically important, and often done incorrectly. Exposure of the sample to differing light intensities will change the sample temperature, and even the act of probing the device in forward bias can lead to resistive heating. Therefore, it is important to mount at least one sensor directly on the substrate as close as possible to the device being measured. A silicon diode temperature sensor is preferable to a thermocouple or other sensors for several reasons. First, it can achieve high accuracy from near absolute zero to 500 K. Secondly, it does not rely on strain changes and therefore can be more reliably affixed to the substrate. Lastly, these sensors often include a flat face, which allows for lower thermal contact resistance with the substrate.

The sample temperature will likely be at least 10–15 K greater than the cold finger temperature due to radiative heat transfer from the lamp and the surrounding atmosphere. This may be limited by using an aperture on the cryostat window or short-pass filter to remove excess infrared lamp light, or good radiation shielding down to the sample position.

## 2.2. Substrate mounting

Proper substrate mounting is necessary for temperature control, and can be challenging as it must withstand cryogenic temperatures and high vacuum (~$10^{-5}$ Torr). For illuminating devices in the substrate configuration there are two straightforward options: (i) a low-volatility cryogenic glue with reasonable thermal conductivity which can help to anchor the substrate as well as eliminate any vacuum gaps between the back of the substrate and the copper stage; or (ii) a silicone-based grease (*e.g.*, Apezon, Dow Corning vacuum grease) can eliminate air gaps and provide temporary weak adhesion while maintaining reasonable thermal conductivity at low temperature. The silicone-based grease is also tolerant to vibration and thermal expansion mismatches because it does not cure like glue. In both cases, however, it is difficult to fully clean the adhesive off the substrate when unmounting after the measurement, and may permanently contaminate the sample.

To electrically isolate the sample (in case of a conductive back contact and front contact) from the copper cold finger mount, we use a thin sapphire plate or a thin layer of kapton/polyimide tape. These materials offer reasonable thermal conductivity for electrical insulators. For the measurement of superstrate-style devices, which require illumination through the substrate, the copper sample mount may be modified with a sapphire window inset into a hole in the copper. Figure 5 gives an overview of some typical $J(V, T, i)$ copper stages used for substrate as well as superstrate illumination.

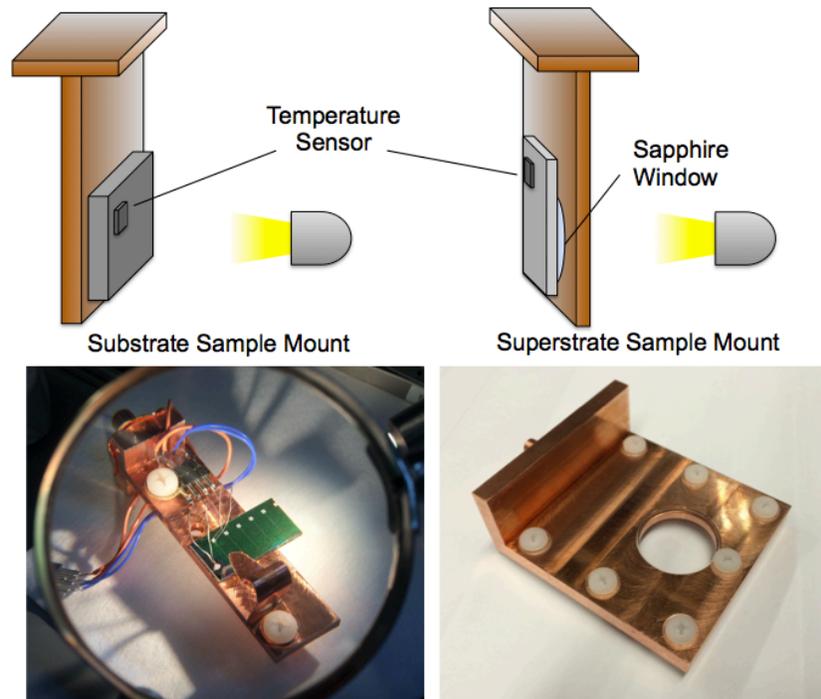

**Figure 5** Custom-machined copper stages for performing $J(V, T, i)$ measurements. Left: a low thermal inertia substrate mount for front-contacted, substrate-style cells. Right: a substrate mount with an inset hole and sapphire window for superstrate illumination. The temperature sensor is mounted in both cases on the same side of the substrate as the electrical contacts.[16]

## 2.3. Electrical contact to the cell

Electrical contact to the cell is ideally made in a 4-point probe configuration, which mitigates the contribution of contact resistance to the measurement by sourcing current and sensing voltage on different pairs of leads. Note that at low temperatures, the contact resistance may increase. There are three strategies for making robust electrical connections. First, mechanical clips made of copper may be used to anchor the substrate and to provide electrical connection at the same time. These may be cut from copper sheets or wires. However, the large mechanical force coupled with substrate vibrations increases the risk of punching through the contact and shunting the device. For soft absorber materials, a more gentle approach is to include a piece of indium foil between the clip and the contact. Device shunting through contacting can be further avoided by using gold wire-bonding between the device contact and a more robust electrical pad outside the sample area. This strategy, however, is more time- and cost-intensive per sample. A lower-cost approach favored in our lab is to use four probe wires (fashioned from thin aluminum or copper wire) attached to the contacts using colloidal silver paste. The paste acts as a weak adhesive and a conduction pathway. It is fast and fairly cheap to apply to each device in a four-point probe configuration. Both the clip-style contacts and the silver-paste wire contacts are visible in Figure 5 (bottom left panel).

In the case that double-tipped probes (or other four-point probe geometries) are used to contact each pad, the contact resistance may be sensed by applying a small voltage between the two double-probe tips using the source-meter before the double-tip probe is lowered onto the top contact. As the double-tip probe is slowly lowered (or as contacts are applied), the voltage reading from the source-meter or the resistance may be monitored. When the probes have made electrical contact, the resistance across the contact changes to a finite value (ideally < 1 Ω), or the source-meter voltage across the device jumps from the noise floor to a steady, finite voltage. This can ensure a gentle contact without applying sufficient pressure to shunt the device.

## 2.4. Illumination control

To provide solar-spectrum illumination, we use a Newport/Oriel LCS-100 Solar Simulator, classified as an ABB simulator with a 1.5" diameter uniform output beam. While under normal operating conditions the lamp outputs AM1.5 intensities, the light intensity may be reduced by using neutral density filters (reflective filters, supplied by Thorlabs). The filters are mounted in a set of two consecutive automated filter wheels. With this setup one can achieve 36 possible illumination intensities over four orders of magnitude in light intensity with only 10 unique filters. The resulting light intensity may be calibrated using a standard silicon photodiode, as long as the photodiode is known to have a linear response to light intensity over the range of interest. The use of neutral density filters is preferred over changing the power input to the lamp, as modifying the power will also change the spectral composition of the light source.

As an alternative to a solar simulator, one may use a monochromatic LED light source. However, to properly evaluate and model the AM1.5 *J-V* behavior, it is important to be able to replicate the true injection conditions that exist at normal operating conditions. This includes not only the generation rate in the absorber layer itself, but also the generation rate in the wide-bandgap window layers and potentially the role of sub-bandgap illumination on trap-filling.

## 3. Cross-sectional electron-beam induced current (EBIC) measurements

Cross-sectional EBIC measurements can reveal current pathways and recombination processes that contribute to current loss including device shunting in a solar cell.[15] However, as with *J(V, T, i)* measurements, it is a time-intensive measurement and requires meticulous tuning of sample preparation, mounting and measurement procedures, the details of which are often not captured in the literature. In the following section, we present detailed experimental methods of our process for EBIC measurements.

Here, we perform cross-sectional EBIC measurement on SnS substrate-style thin film solar cells (see Figure 1c for the device stack and Figure 2 for an image of an identical device substrate) as a proof-of-concept. EBIC measurements are performed in a dual-beam focused ion beam-secondary electron microscope (Helios NanoLab 600, FEI), equipped with an EBIC system (Point Electronic DISS5) under high vacuum conditions at room temperature.

### 3.1. Sample preparation

For sample preparation, we cleave the device sample (of dimensions 9 mm × 2.75 mm) midway along the long dimension of the device, leaving a 2.75 mm length of the device cross-section exposed to air.

Due to the polycrystalline nature of SnS thin films, the SnS layer generally cleaves along the grain boundaries, resulting in a relatively rough topology at the cross-section. The device stack in Figure 1c depicts a cross-sectional micrograph of an SnS thin film polycrystalline morphology. Additional scanning electron micrographs of SnS thin films have been published elsewhere.[19,27,28]

To flatten the topology at the cross-section, we developed a four-step polishing process, using a $Ga^+$ focused ion beam. The polishing process is time-intensive; however this process has been shown to be gentle enough to polish soft material cross-sections reliably without risking device shunting, which is critical for cross-sectional EBIC measurements. We polish a 20 μm-wide segment of the device cross-section.

The device stack before and after polishing are shown in Figure 6a and 6b, respectively. First, a platinum pad (20 μm x 2 μm x ~1 μm thick) is deposited to protect the top surface of the device and to provide planarization that prevents inhomogeneous milling due to the "curtaining effect"[29] (Figure 6a). The platinum is deposited *in situ* using the built-in gas injection system of the dual-beam system using the focused ion beam as the excitation source to promote localized decomposition of the precursor. The first milling step uses an ion beam with beam current of 2.7 nA and an accelerating voltage of 30 kV over a rectangular area (20 μm x 500 nm) near the edge of the cleaved cross-section. The ion beam is tilted at an angle of 2.5 degrees off the top surface normal of the device, toward the exposed cross-section face, compensating for the reduction in sputtering yield as a function of depth,[30] leaving a polished cross-section face that is normal to the top surface. The first milling step is completed once the ion beam has milled from the cleaved edge into the platinum-protected region, as monitored by the scanning electron microscope. The second milling step uses a lower beam current of 91 pA while the acceleration voltage is kept constant at 30 kV. This milling process is carried out over the same 20 μm-wide segment, at a 2 degree tilt relative to the normal in a line-by-line fashion to polish the cross-section more gently. The third ion milling step uses a beam current of 150 pA and 5 kV acceleration voltage at a 3 degree tilt. The purpose of the final milling step is to reduce the thickness of the amorphous region at the milled face that occurs due to damage from the 30 kV Ga$^+$ ions.[31] This multi-step milling procedure is used to provide a surface that mitigates surface damage while also enabling a reasonable throughput.

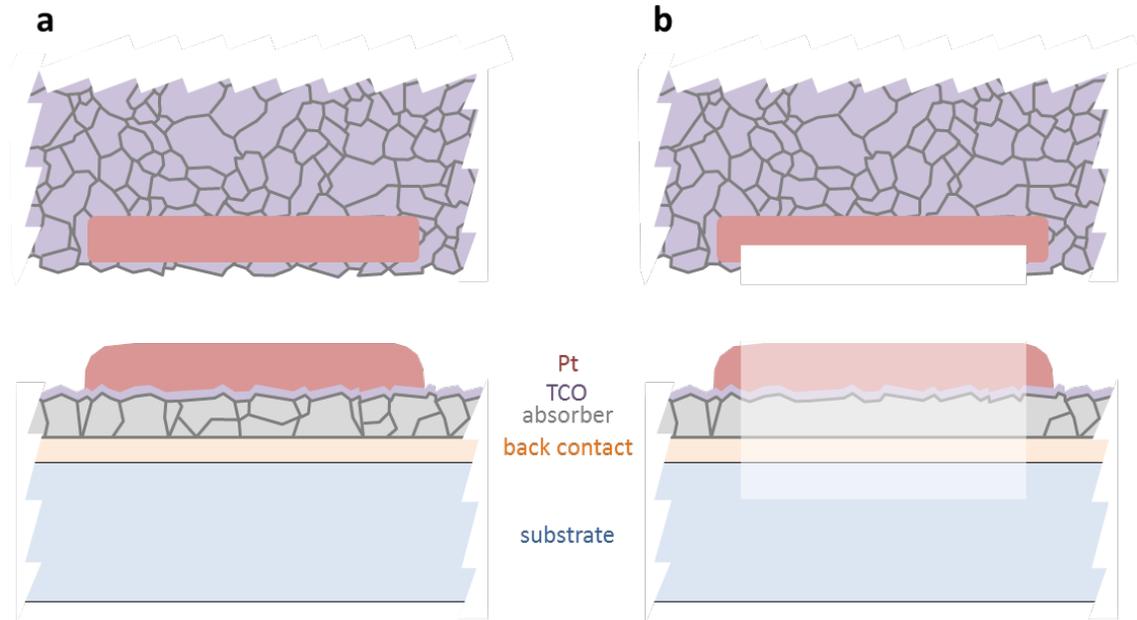

**Figure 6** Diagram of polishing process from plan view (top) and side view (bottom). (a) Device stack just before polishing procedure. A platinum bar (red) is deposited on top of the freshly cleaved device stack near the cross-section edge to protect the TCO layer in the region of interest and to provide planarization for the milling steps. The film generally cleaves along grain boundaries so that grain boundaries are clearly visible in the cross-section. (b) Device stack after ion-milling process. Ion milling of the device stack results in a smooth cross-section (rectangular region) on which to measure EBIC. Grain boundaries are obscured in the ion-milled region because of the smooth topology.

### 3.2. Sample mounting

To mount the device with polished cross-section onto a custom EBIC sample stage, we unload the sample from the microscope for about 10 – 20 minutes. Depending on the material, it may be advantageous to leave the sample exposed to ambient air for a longer period of time to form a passivating oxide layer on the newly polished cross-section. It is important to mount the sample in a dust-free environment and to avoid touching the device cross-section with any other object, as dust particles adhered to the cross-section will obscure EBIC results.

The sample is affixed to the stage using double-sided copper tape. We expose the metal back contact of the substrate-style device by mechanically exfoliating the top device layers with a razor blade next to the active device area. A 2 mm x 1 mm x 0.5 mm indium bar is carefully placed onto the metal top contact pad of the device to serve as a soft mechanical buffer between the metal contact pad and the electrical probe of the EBIC stage. It is important that the indium is gently placed onto the contact pad accurately, and not slid once mechanical contact is made. Sliding the indium bar on the device may scratch the device and cause shunts, despite the softness of the indium. The indium mechanical buffer reduces the risk of device shunting from excessive pressure and sliding of the electrical probe of the EBIC stage, similar to the *J(V, T, i)* setup discussed in Section 2.

The two electrical probes consist of copper-beryllium wire spot-welded to a stainless steel washer, which is fixed by a screw into the base of the stage. One of the probes is placed on the exposed metal back contact of the device and the other is placed on the indium bar on the top contact of the device. Figure 7 shows a photograph of the sample mounted onto the EBIC sample stage.

The mounted sample is then placed back into the same dual-beam microscope that is equipped with an EBIC system. Once mounted to the stage, the electrical feed-through leads of the EBIC system are connected to the sample mount. Before pumping down the dual-beam microscope, a current-voltage measurement is performed (typically from -0.5 to 0.5 V) as a test of proper electrical contact and rectification. Once electrical contact and rectification are confirmed, the dual-beam microscope is pumped down. If the current-voltage curve exhibits an ohmic characteristic, the device was likely shunted in the contacting process, in which case another device must be prepared.

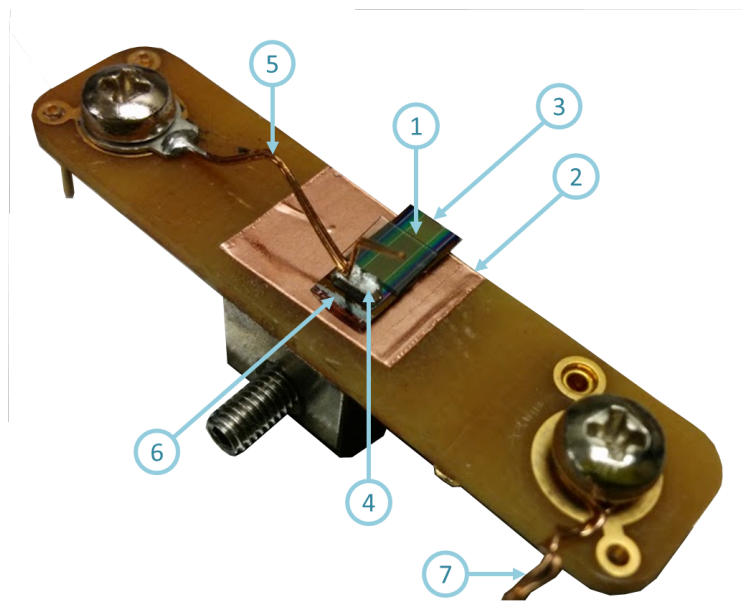

**Figure 7** Photograph of sample mounted onto the EBIC sample holder. The device sample (1) is fixed to the holder using double-sided copper tape (2), with the exposed device cross-section (3) overhanging slightly off of the sample holder edge. An indium bar (4) is placed on the top contact pad of the device, and the electrical probe (5) is placed on top of the indium bar. An inactive area of the device is abraded with a razor blade to expose the bottom metal contact of the device stack (6), onto which the second electrical probe (7) would be placed (the second contact is not yet formed in the photograph for clarity).

## 3.3. Measurement

The electron accelerating voltage should be chosen based on the desired electron excitation volume inside the absorber layer material. To first order the electron range in a given material varies inversely with the density and has a power law dependence on the accelerating voltage with an exponent of ~1.5-1.7.[32] The electron beam current for our measurements is typically 86 pA to balance the tradeoff between achieving a high signal-to-noise ratio and moderating the injection level (low energy electron beams, <10 keV, tend to lead to high-level injection as the excitation volume is very small since the electron range is typically <1 μm in this regime). The CASINO software package is a freely available Monte Carlo simulation tool for estimating the generation volume for a variety of sample geometries.[33]

To optimize EBIC image acquisition it is necessary to maximize the intensity of the analog signal to be digitized without clipping. This provides the highest current resolution without sacrificing data. This is achieved by setting three parameters for EBIC system: gain, offset, and contrast.

The gain value is set first, which determines the transimpedance amplification of the pre-amplifier. To enable the collection of a variety of current signals the transimpedance gain can be varied over seven orders of magnitude, from $10^3$-$10^{10}$ V/A. The gain value is set by monitoring a live waveform monitor in the EBIC acquisition software. A suitable gain is the largest value (in decade increments) that does not saturate the digital-to-analog converter (DAC). Once the gain is set, the offset value must be determined. With the contrast minimized, the desired offset value will center the output in the measurement range of the DAC. Finally, the contrast should be maximized such that the waveform is within the measurement range of the DAC. Because the offset is determined before the contrast level this ensures that the signal will be amplified about the middle of the DAC range, leading to maximum current resolution before digitization without clipping.

## 4. Summary and conclusions

In this article we shared best practices and experimental experiences on thin film solar cell designs, sample preparation and sample mounting for *J(V, T, i)* and EBIC on the example of tin monosulfide, a promising emerging thin film absorber candidate material. We found the use of soft probe tips, mechanical buffers or wire contacting helpful for shunt prevention during *J(V, T, i)* measurements (see section 2.3. and 3.2.). To reliably mount the substrate to the copper cold finger in the cryostat setup, we found the use of silicone-based grease to be most helpful (see section 2.2.). In section 3.1., we developed a reliable four-step cross-section polishing process for SnS sample preparation for cross-sectional EBIC. In section 3.2. we demonstrated how the use of an indium bar can be useful to prevent shunting during EBIC measurements. Finally, Section 3.3 describes the process of optimizing electrical parameters for high-quality EBIC acquisition.

Moving forward, the discussed techniques and best practices should be considered when exploring novel thin film solar cells material systems for high-efficiency PV devices to avoid "false negatives".